\PassOptionsToPackage{unicode}{hyperref}
\PassOptionsToPackage{hyphens}{url}
\PassOptionsToPackage{dvipsnames,svgnames,x11names}{xcolor}
\documentclass[
  12pt]{article}

\usepackage{amsmath,amssymb}
\allowdisplaybreaks[3]
\usepackage{iftex}
\ifPDFTeX
  \usepackage[T1]{fontenc}
  \usepackage[utf8]{inputenc}
  \usepackage{textcomp} 
\else 
  \usepackage{unicode-math}
  \defaultfontfeatures{Scale=MatchLowercase}
  \defaultfontfeatures[\rmfamily]{Ligatures=TeX,Scale=1}
\fi
\usepackage{lmodern}
\ifPDFTeX\else  
\fi
\IfFileExists{upquote.sty}{\usepackage{upquote}}{}
\IfFileExists{microtype.sty}{
  \usepackage[]{microtype}
  \UseMicrotypeSet[protrusion]{basicmath} 
}{}
\makeatletter
\@ifundefined{KOMAClassName}{
  \IfFileExists{parskip.sty}{%
    \usepackage{parskip}
  }{
    \setlength{\parindent}{0pt}
    \setlength{\parskip}{6pt plus 2pt minus 1pt}}
}{
  \KOMAoptions{parskip=half}}
\makeatother
\usepackage{xcolor}
\makeatletter
\ifx\paragraph\undefined\else
  \let\oldparagraph\paragraph
  \renewcommand{\paragraph}{
    \@ifstar
      \xxxParagraphStar
      \xxxParagraphNoStar
  }
  \newcommand{\xxxParagraphStar}[1]{\oldparagraph*{#1}\mbox{}}
  \newcommand{\xxxParagraphNoStar}[1]{\oldparagraph{#1}\mbox{}}
\fi
\ifx\subparagraph\undefined\else
  \let\oldsubparagraph\subparagraph
  \renewcommand{\subparagraph}{
    \@ifstar
      \xxxSubParagraphStar
      \xxxSubParagraphNoStar
  }
  \newcommand{\xxxSubParagraphStar}[1]{\oldsubparagraph*{#1}\mbox{}}
  \newcommand{\xxxSubParagraphNoStar}[1]{\oldsubparagraph{#1}\mbox{}}
\fi
\makeatother

\usepackage{longtable,booktabs,array}
\usepackage{calc} 
\usepackage{etoolbox}
\makeatletter
\patchcmd\longtable{\par}{\if@noskipsec\mbox{}\fi\par}{}{}
\makeatother
\IfFileExists{footnotehyper.sty}{\usepackage{footnotehyper}}{\usepackage{footnote}}
\makesavenoteenv{longtable}
\usepackage{graphicx}
\makeatletter
\def\maxwidth{\ifdim\Gin@nat@width>\linewidth\linewidth\else\Gin@nat@width\fi}
\def\maxheight{\ifdim\Gin@nat@height>\textheight\textheight\else\Gin@nat@height\fi}
\makeatother
\setkeys{Gin}{width=\maxwidth,height=\maxheight,keepaspectratio}
\makeatletter
\def\fps@figure{htbp}
\makeatother

\makeatletter
\@ifpackageloaded{caption}{}{\usepackage{caption}}
\AtBeginDocument{%
\ifdefined\contentsname
  \renewcommand*\contentsname{Table of contents}
\else
  \newcommand\contentsname{Table of contents}
\fi
\ifdefined\listfigurename
  \renewcommand*\listfigurename{List of Figures}
\else
  \newcommand\listfigurename{List of Figures}
\fi
\ifdefined\listtablename
  \renewcommand*\listtablename{List of Tables}
\else
  \newcommand\listtablename{List of Tables}
\fi
\ifdefined\figurename
  \renewcommand*\figurename{Figure}
\else
  \newcommand\figurename{Figure}
\fi
\ifdefined\tablename
  \renewcommand*\tablename{Table}
\else
  \newcommand\tablename{Table}
\fi
}
\@ifpackageloaded{float}{}{\usepackage{float}}
\floatstyle{ruled}
\@ifundefined{c@chapter}{\newfloat{codelisting}{h}{lop}}{\newfloat{codelisting}{h}{lop}[chapter]}
\floatname{codelisting}{Listing}

\makeatother
\makeatletter
\@ifpackageloaded{caption}{}{\usepackage{caption}}
\@ifpackageloaded{subcaption}{}{\usepackage{subcaption}}
\makeatother
\AtBeginEnvironment{tabular}{%
  \footnotesize
  \setlength{\tabcolsep}{3.5pt}%
  \setlength{\aboverulesep}{0.25ex}%
  \setlength{\belowrulesep}{0.25ex}%
  \setlength{\defaultaddspace}{0.25ex}%
}
\AtBeginEnvironment{longtable}{%
  \footnotesize
  \setlength{\tabcolsep}{3.5pt}%
  \setlength{\aboverulesep}{0.25ex}%
  \setlength{\belowrulesep}{0.25ex}%
  \setlength{\defaultaddspace}{0.25ex}%
}
\ifLuaTeX
  \usepackage{selnolig}  
\fi
\usepackage[authoryear,round]{natbib}
\bibpunct{(}{)}{;}{a}{}{,}
\usepackage{bookmark}

\IfFileExists{xurl.sty}{\usepackage{xurl}}{} 
\hypersetup{
  pdftitle={Title},
  pdfauthor={Author 1; Author 2},
  pdfkeywords={3 to 6 keywords, that do not appear in the title},
  colorlinks=true,
  linkcolor={blue},
  filecolor={Maroon},
  citecolor={Blue},
  urlcolor={Blue},
  pdfcreator={LaTeX via pandoc}}

\newcommand{\anon}{1}

\begin{document}

\newtheorem{theorem}{Theorem}
\newtheorem{proposition}{Proposition}
\newtheorem{corollary}{Corollary}
\newtheorem{lemma}{Lemma}
\newtheorem{remark}{Remark}
\newtheorem{definition}{Definition}
\newtheorem{assumption}{Assumption}
\newenvironment{proof}[1][Proof]{\par\noindent\textit{#1.} }{\hfill$\square$\par}
\providecommand{\FloatBarrier}{}
\newcommand{\argmin}{\mathop{\mathrm{argmin}}\limits}
\floatstyle{ruled}
\newfloat{algorithm}{htbp}{loa}
\floatname{algorithm}{Algorithm}
\newcommand{\SetCustomAlgoRuledWidth}[1]{}
\long\def\If#1#2{\par\noindent\textbf{If} #1,\par #2\par}
\long\def\ElseIf#1#2{\par\noindent\textbf{Else if} #1,\par #2\par}
\hypersetup{
  pdftitle={Transporting Trial Evidence Under Posterior Drift and Possible Hidden Confounding},
  pdfauthor={Bosen Cui; Xilin Mao; Yuhong Yang},
  pdfkeywords={Causal inference; Data integration;  Observational analysis; Robust estimation; Transportability}
}

\def\spacingset#1{\renewcommand{\baselinestretch}%
{#1}\small\normalsize} \spacingset{1}


\if1\anon
{
  \title{\bf Transporting Trial Evidence Under Posterior Drift and Possible Hidden Confounding}
  \author{Xilin Mao$^{\dagger}$\\
    Qiuzhen College, Tsinghua University\\
    \\
    Bosen Cui$^{\dagger}$\\
    Yau Mathematical Sciences Center, Tsinghua University\\
    \\
    Yuhong Yang$^*$\\
    Yau Mathematical Sciences Center, Tsinghua University\\
    Beijing Institute of Mathematical Sciences and Applications}
  \maketitle
  \date{}
  \begin{center} \begin{minipage}{0.85\textwidth} \footnotesize $^\dagger$ Xilin Mao and Bosen Cui contributed equally to this work.\\ $^*$ Corresponding author. Email: yyangsc@tsinghua.edu.cn \end{minipage} \end{center}
} \fi

\if0\anon
{
  \bigskip
  \bigskip
  \bigskip
  \begin{center}
    {\LARGE\bf Transporting Trial Evidence Under Posterior Drift and Possible Hidden Confounding}
\end{center}
  \medskip
} \fi

\bigskip
\begin{abstract}
Randomized trials provide internally valid treatment-effect evidence, but trial participants may not represent the target population. In contrast, observational studies are often closer to the target population, but their treatment assignment may be affected by possible hidden confounding. We develop a robust posterior-drift framework for estimating the average treatment effect in an observational target population when exact conditional-effect transportability may fail. The framework represents observational conditional potential-outcome regressions as their randomized-trial counterparts plus source-specific drifts. The randomized trial serves as an internally valid anchor, while the observational study supplies the target covariate distribution and partial information about the target causal contrast. To account for possible hidden confounding, we consider a Rosenbaum-type uncertainty set induced by a sensitivity parameter on the generalized propensity score and estimate the drift through a minimax worst-case risk criterion. We derive efficiency results in auxiliary regimes, establish uniform concentration and near-optimality guarantees for the minimax estimator, and handle general parametric and smooth nonparametric drift classes. Simulations and an ACTG 175--WIHS application show that the proposed analysis yields more cautious and interpretable target-population effect estimates than exact-transportability analyses.
\end{abstract}

\noindent%
\vfill

\newpage
\spacingset{1.8} 

\section{Introduction}

Randomized controlled trials (RCTs) and observational studies (OSs) provide
complementary sources of evidence for causal inference. Randomization protects
RCTs against confounding and gives them strong internal validity, but trial
participants are often highly selected and may not represent the population in
which treatment decisions are ultimately made \citep{Cole2010generalizing}.
Observational studies, by contrast, are often larger, less costly to conduct,
and more representative of routine practice, but treatment assignment is not
randomized and may therefore be affected by measured or unmeasured confounding.
A central statistical problem is therefore how to transport internally valid
trial evidence to an observational target population. Existing work on trial
generalizability and transportability has developed weighting, outcome-modeling,
and doubly robust estimators for both nested and non-nested designs
\citep{Cole2010generalizing,Dahabreh2019ontherelation,Dahabreh2020Extendinginferences};
see also \citet{Degtiar2023generalizabilityreview},
\citet{Colnet2024causalcombiningsurvey}, and \citet{Dahabreh2024review} for
recent reviews.

Most transportability methods rely, explicitly or implicitly, on a stability
condition linking the trial and target populations. A common version assumes
that the conditional average treatment effect (CATE) in the RCT coincides with
the CATE in the OS after adjusting for observed covariates. Under this condition,
one may estimate the trial CATE and average it over the OS covariate distribution
to recover the target-population average treatment effect (ATE). However, exact
conditional-effect transportability can be difficult to justify in practice.
Eligibility criteria, adherence, monitoring intensity, clinical practice, and
latent health characteristics may affect both trial participation and treatment
response. Recent works have therefore considered relaxations based on bias functions
or posterior drift, allowing conditional outcome regressions to differ across
data sources \citep{Dahabreh2023sensitivityanalysis,wu2024comparativeanalysisaveragetreatment}.
Related source--target discrepancies have also been studied in the transfer-learning
literature under posterior-drift or related regression-shift models
\citep{maity2021linearadjustmentbasedapproach,tonycai2021nonparametricclassification,
cai2024transferlearningnonparametricregression}. Another related line of work
combines experimental and observational evidence, including methods that borrow
observational or external-control information to improve randomized-trial analyses
\citep{Yang2023Elasticintegrative,Gao2025improvingrandomizedcontrolled,
yang2026improvingtreatmenteffectestimation} and broader multi-source methods for
target-effect estimation \citep{yang2020combining,Han2025Federatedaptive}.
Together, these works address important forms of data-source heterogeneity,
but they do not directly handle estimation of an observational target-population
ATE when cross-source posterior drift is unknown and the OS treatment assignment
may be hidden-confounded. This gap is practically important because many real-world
target populations are represented only through observational data, precisely where
residual confounding and violations of exact transportability are most difficult to rule out.

In this paper, we develop a posterior-drift framework for estimating the ATE in
an observational target population when exact transportability may fail and the
observational treatment assignment may be affected by hidden confounding. We
write the OS conditional potential-outcomes as the corresponding RCT
conditional potential-outcomes plus source-specific drifts. The RCT therefore serves as an
internally valid anchor for the baseline treatment-response surface, while the OS
supplies the target covariate distribution and partial information about the
target causal contrast. Because hidden confounding prevents direct identification
of the OS CATE from the OS data, we place a Rosenbaum-type
sensitivity model on the generalized propensity score and estimate the drift by
a minimax criterion over the resulting uncertainty set. 
This minimax component is related to recent robust and
marginal-sensitivity approaches to hidden confounding
\citep{Kallus2021minimaxpolicyconfounding,
sahoo2024learningbiasedsample,Dorn2025doublyvalid}, but our use of a randomized trial as an anchor
for cross-source drift estimation is different from single-source robust methods for hidden confounding, robust policy learning, or prediction problems.

Our contributions are threefold. First, we formulate target-population ATE
estimation under simultaneous posterior drift and possible hidden confounding,
separating this problem from standard trial transportability, external-control
borrowing, and single-source robust analyses of hidden confounding. Compared with approaches that
impose exact equality or a known functional relationship across sources
\citep{Li2020improvingefficiency,wu2024comparativeanalysisaveragetreatment}, our
framework treats the cross-source discrepancy as an unknown object to be estimated
and calibrated. Second, we develop a minimax posterior-drift estimator that
combines RCT anchoring, OS target information, and uncertainty-set calibration. We
establish uniform concentration and near-optimality guarantees for the resulting
worst-case risk, thereby connecting the finite-sample optimization problem to
the population-level minimax target. Third, although a linear working drift
class is used to present the core construction and facilitate implementation,
the framework is not tied to linearity. We expand the minimax construction
to handle finite-dimensional parametric drift classes and to smooth nonparametric drift
classes as well through sieve approximation.

Numerically, our method is evaluated in simulation studies and in a non-nested trial--cohort
application that transports evidence from ACTG 175, an HIV randomized trial, to
the Women's Interagency HIV Study (WIHS) cohort
\citep{HammerKatzenstein1996ACTG,Adimora2018WIHS}. In the real data application, 
after finding incompatibility between the conditional treatment-effect contrasts identified from the two sources, our robust
posterior-drift analysis yields more cautious and interpretable
target-population effect estimates than analyses that impose exact
transportability. These findings illustrate the practical value of replacing
brittle equality assumptions by an explicit drift-and-robustness framework when
integrating experimental and observational evidence.

The remainder of the article is organized as follows. Section~2 introduces the
setup and target estimand. Section~3 presents two auxiliary linear posterior-drift regimes: one with known
drift and one with OS unconfoundedness. Section~4 develops
the main minimax robust procedure for unknown drift under hidden
confounding and extends the framework to parametric and nonparametric drift
classes. Section~5 reports simulation results. Section~6 presents the ACTG
175--WIHS application, and Section~7 concludes.

\section{Setup}\label{sec: setup}
We consider a setting where each unit is characterized by pre-treatment covariates $X\in\mathcal{X}\subset \mathbb{R}^d$, a binary treatment
$A\in\{0,1\}$, and an outcome $Y\in\mathbb{R}$. Under the potential-outcomes framework
\citep{neyman1923application,rubin1974estimating}, each unit has two potential outcomes
$Y(0)$ and $Y(1)$. We proceed under the  Stable Unit Treatment Value Assumption (SUTVA) and the consistency assumption, so the observed outcome satisfies
\(
Y=Y(A)=AY(1)+(1-A)Y(0)
\).

Our analysis combines two data sources: an RCT and an OS. Let $G\in\{0,1\}$
indicate data source, where $G=1$ denotes the RCT and $G=0$ denotes the OS. Following the superpopulation formulations commonly used in trial transportability studies \citep{Dahabreh2019ontherelation,wu2024comparativeanalysisaveragetreatment},  
we assume
\(
\{(X_i,A_i,Y_i(0),Y_i(1),G_i)\}_{i=1}^{N_{\rm tot}}
\)
are i.i.d. draws from a joint distribution $\mathbb P$. Here $N_{\rm tot}$
denotes the total number of available observations,
and the observed data are
\(
\{(X_i,A_i,Y_i,G_i)\}_{i=1}^{N_{\rm tot}}.
\)

Let
$q:=\mathbb{P}(G=1)$ denote the probability of being included in the RCT.
Our central estimand is the ATE in the OS population,
\(
\tau:=\mathbb E\{Y(1)-Y(0)\mid G=0\},
\)
a key target quantity in trial transportability problems
\citep{Cole2010generalizing, Dahabreh2024review}.
For $g\in\{0,1\}$, define the source-specific propensity score
\(
e_g(x):=\mathbb{P}(A=1\mid X=x,G=g),
\)
and define the sampling score
\(
\pi(x):=\mathbb{P}(G=1\mid X=x).
\)
For $a\in\{0,1\}$, define
\(\mu_a(x):=\mathbb{E}(Y(a)\mid X=x,G=1)\), \(\tilde{\mu}_a(x):=\mathbb{E}\{Y(a)\mid X=x,G=0\}\)
as the conditional mean potential outcomes in the RCT and OS populations, respectively. We further define
\(
\Delta(x):=\mu_1(x)-\mu_0(x)
\),
\(
\tilde{\Delta}(x)
:=
\tilde{\mu}_1(x)-\tilde{\mu}_0(x)
\), 
which are the CATE functions in the RCT and OS populations, respectively.

\begin{remark}
The above two-source superpopulation formulation is compatible with both nested trial designs and non-nested composite-data designs. In a nested trial design, $G$ is interpreted as trial participation within a cohort or sample drawn from the target population. In a non-nested composite-data design, $G$ instead indicates whether an observation comes from a completed randomized trial or from a separately obtained observational sample representing the target population. The latter interpretation is particularly relevant for applications that append trial data to external cohort, registry, or electronic-health-record data. The sampling score $\pi(x)$ should therefore be interpreted according to the underlying design: as a trial-participation probability in nested designs and as a source-membership probability in non-nested composite-data designs. See \citet{Dahabreh2020Studydesignsfor} and \citet{Dahabreh2020Extendinginferences} for a systematic discussion of nested and non-nested trial designs.
\end{remark}

With the notation in place, we can state the trial-validity condition that anchors all subsequent analyses.
\begin{assumption}[Internal validity of RCT]\label{assumption: unconfounding RCT} (i)
    $A\perp\!\!\!\perp \{Y(0), Y(1)\}|X,G=1$; (ii) $0<e_1(x) <1$ for all $x \in \mathcal{X}$. 
\end{assumption}
Assumption \ref{assumption: unconfounding RCT} (i) reflects the randomized nature of the RCT, which eliminates confounding and ensures internal validity. Assumption \ref{assumption: unconfounding RCT}(ii) guarantees that each subject in the RCT has a positive probability of receiving each version of treatment. In most cases, $e_1(x)$ is known in RCT design.

A commonly used transportability condition in the two-dataset setting is the following one \citep{Pearl2011tansportability,Dahabreh2020Studydesignsfor,Li2020improvingefficiency}:
\[
    \mathbb{E}(Y(1)-Y(0) \mid X, G=1) = \mathbb{E}(Y(1)-Y(0) \mid X, G=0).
\]
This assumption requires that the CATE be transportable in the RCT and OS. Under
this condition, target parameters in the OS that depend on \(\Delta(X)\) can in
principle be identified by substituting a CATE estimator trained in the RCT,
which greatly simplifies inference. However, in many applications this assumption
is implausible: differences in enrollment criteria, clinical practice, or
measurement protocols may induce systematic discrepancies between the two CATE
functions. Rather than imposing this exact CATE transportability as a global
assumption, we next
introduce posterior-drift formulations that allow the OS conditional
potential-outcome expectations to differ from their RCT counterparts.

\section{Auxiliary regimes: known drift and OS unconfoundedness}\label{sec:linear parameters known}

This section introduces the linear posterior-drift model and studies two auxiliary
regimes under this model. The first is an oracle regime in which the linear drift coefficient is known; the second is an OS-unconfounded regime in which the drift coefficient can
be estimated directly from the difference between trial and observational CATE
regressions.

\begin{assumption}[Linear posterior drift]\label{assumption: linear posterior drift}
$\tilde{\mu}_0(X) = \mu_0(X) + \alpha^\top X$ and $\tilde{\mu}_1(X) = \mu_1(X) + \beta^\top X$, where $\alpha$ and $\beta$ are fixed parameter vectors.
\end{assumption}

Under Assumption~\ref{assumption: linear posterior drift}, the differences between the RCT
and OS conditional potential-outcome expectations are linear in $X$ for both treatment
arms. 
Writing
\(\gamma=\beta-\alpha\), the OS CATE satisfies
\[
\tilde{\Delta}(X)
=
\Delta(X)+\gamma^\top X.
\]
Thus, the cross-source CATE discrepancy is reduced to the finite-dimensional
drift coefficient \(\gamma\). 

\begin{remark}
    This linear specification is used to present the core construction and efficiency
comparisons; more flexible parametric and nonparametric drift classes are studied
in Section~\ref{sec: hidden confounding, linear parameters unknown}.

\end{remark}

\subsection{Known linear drift and efficiency gain}

Because we do not impose a parametric error distribution for the conditional
outcome regressions, we work in a semiparametric framework rather than relying
on a fully parametric likelihood. Following a common practice in modern causal
inference \citep{Robins1994EstimationofRegressionCoefficients,Tsiatis2008covariateadjustment,Tan2010Boundedefficientdoublyrobust,Chan2016Globallyefficientnonparametric}, we combine the
information from the RCT and OS datasets to construct estimators for the ATE in the OS population.

Our estimation strategy is built around the efficient influence function (EIF) for the target parameter $\tau$. We first derive the EIF to characterize the semiparametric efficiency bound and to identify the correction terms needed to combine trial and observational information. Guided by this representation, we then construct efficient estimators for $\tau$ by plugging in estimates of the nuisance components. This formulation naturally accommodates machine-learning or other data-adaptive nuisance estimators while preserving valid statistical inference under the stated conditions.

Within the known-drift regime, we consider two settings. The first one 
uses only the RCT validity condition and the linear posterior-drift model. The
second one additionally assumes unconfoundedness of OS data. We state
this extra assumption before presenting the EIFs.

\begin{assumption}[Unconfoundedness and positivity of OS]\label{assumption: unconfounding OS}
    $A\perp\!\!\!\perp \{Y(0), Y(1)\}|X,G=0$, and $0<e_0(x) <1$ for all $x$.
\end{assumption}

Under Assumption~\ref{assumption: unconfounding OS}, the OS-population ATE $\tau$ is nonparametrically identified from the OS alone. In this auxiliary regime, Assumption~\ref{assumption: linear posterior drift} is therefore not required for identification; rather, it provides cross-source structure linking the RCT and OS conditional outcome regressions. Assumption~\ref{assumption: unconfounding OS} is used only in the auxiliary analyses of Section 3 and is not imposed in the hidden-confounding analysis of Section 4 or thereafter.

\begin{theorem}[EIFs of $\tau$]\label{EIF}
Suppose Assumptions \ref{assumption: unconfounding RCT} and \ref{assumption: linear posterior drift} hold, and take $\tau := \mathbb{E}[Y(1)-Y(0)|G=0]$ as the estimand. We have the following results: 
    \begin{enumerate}\renewcommand{\labelenumi}{(\alph{enumi})}
        \item the EIF is given as 
        \begin{align*}
        \phi_I &= \frac{G}{1-q} \frac{1-\pi(X)}{\pi(X)} \left[ \frac{A\{Y - \mu_1(X)\}}{e_1(X)} - \frac{(1-A)\{Y - \mu_0(X)\}}{1-e_1(X)} \right] \\
        &\quad + \frac{1-G}{1-q} [\mu_1(X)-\mu_0(X)+\gamma^{\top}X - \tau].
        \end{align*}
        \item Moreover, if Assumption \ref{assumption: unconfounding OS} holds, let
        \[
        r_1(X) = \frac{ \text{Var}(Y(1)|X, G=1)}{\text{Var}(Y(1)|X, G=0)}, \quad r_0(X) = \frac{ \text{Var}(Y(0)|X, G=1)}{\text{Var}(Y(0)|X, G=0)},
        \]
        and define
        \(\omega_1(X) = \pi(X) e_1(X) + (1-\pi(X)) e_0(X) r_1(X)\), 
        \(\omega_0(X) = \pi(X) (1-e_1(X)) + (1-\pi(X)) (1-e_0(X)) r_0(X)\),
        then the EIF is given as 
        \begin{align*}
        \phi_{II} &= \frac{G(1-\pi(X))}{1-q} \left[ \frac{ A\{Y-\mu_1(X)\}}{\omega_1(X)} - \frac{(1-A)\{Y-\mu_0(X)\}}{\omega_0(X)} \right] \\
        &\quad + \frac{(1-G)(1-\pi(X))}{1-q} \left[ \frac{r_1(X) \cdot A\{Y-\tilde{\mu}_1(X)\}}{\omega_1(X)} - \frac{r_0(X) \cdot (1-A)\{Y-\tilde{\mu}_0(X)\}}{\omega_0(X)} \right] \\
        &\quad + \frac{1-G}{1-q} [\mu_1(X)-\mu_0(X)+\gamma^{\top}X - \tau].
        \end{align*}
    \end{enumerate}
    The corresponding efficiency bounds are $V^*_I = \mathbb{E}(\phi_I^2)$, $V^*_{II} = \mathbb{E}(\phi_{II}^2)$.
\end{theorem}

Theorem~\ref{EIF}(b) characterizes the efficiency bound when both the RCT and OS are used under this cross-source structure. Since $\tau$ is already identified from the OS alone under Assumption~\ref{assumption: unconfounding OS}, a natural question is whether incorporating the RCT provides an efficiency advantage. The following proposition answers this question by comparing the combined-data efficiency bound $V^*_{II}$ with the OS-only efficiency bound $V^*$.

\begin{proposition}[Efficiency gain]
    Suppose Assumptions \ref{assumption: unconfounding RCT}, \ref{assumption: linear posterior drift}, \ref{assumption: unconfounding OS} hold. Let $V^*$ denote the semiparametric efficiency bound for estimating $\tau$ when utilizing the OS data (G=0) in isolation. By incorporating the RCT data (G=1) under the linear posterior drift framework, the resulting efficiency bound $V_{II}^*$ satisfies $V_{II}^* \leq V^*$, where the explicit efficiency gain is:
    \begin{align*}
    V^*-V^*_{II}
    &= \mathbb{E}\left[ \frac{(1-\pi(X))\pi(X)e_1(X)}{(1-q)^2} \frac{\text{Var}(Y(1)|X, G=0)}{\omega_1(X)e_0(X)} \right. \\
    &\quad \left. + \frac{(1-\pi(X))\pi(X)(1-e_1(X))}{(1-q)^2} \frac{\text{Var}(Y(0)|X, G=0)}{\omega_0(X)(1-e_0(X))} \right] \geq 0.
    \end{align*}
\end{proposition} 

Having established the necessary efficiency foundations, we now construct the
corresponding M-estimators. To separate nuisance estimation from ATE estimation, we use a source-stratified two-way split. Specifically, an analysis
sample with \(2n\) RCT observations and \(2\tilde n\) OS observations is split
into a nuisance-training fold and an ATE-estimation fold, each containing \(n\)
RCT observations and \(\tilde n\) OS observations. Thus each fold has size
\(m=n+\tilde n\), with the same RCT/OS proportion as the full analysis sample.
Let \(\mathbb P_m\) denote the empirical average over the ATE-estimation fold,
while all nuisance functions below are estimated on the nuisance-training fold.

If only Assumptions \ref{assumption: unconfounding RCT} and \ref{assumption: linear posterior drift} hold, we have 
\begin{equation}\label{equation: estimator with known linear drift, A1+A2}
    \begin{aligned}
        \hat{\tau} = &\mathbb{P}_m\left( \frac{G}{1-q} \frac{1-\hat{\pi}(X)}{\hat{\pi}(X)} \frac{A\{Y-\hat{\mu}_1(X)\}}{\hat{e}_1(X)} \right.
- \frac{G}{1-q} \frac{1-\hat{\pi}(X)}{\hat{\pi}(X)} \frac{(1-A)\{Y-\hat{\mu}_0(X)\}}{1-\hat{e}_1(X)}\\
    & \left. + \frac{1-G}{1-q} [\hat{\mu}_1(X)-\hat{\mu}_0(X)+\gamma^{\top}X] \right).
    \end{aligned}
\end{equation}
Furthermore, if Assumption \ref{assumption: unconfounding OS} holds as well, then 
\begin{equation}\label{equation: estimator with known linear drift, A1+A2+A3}
    \begin{aligned}
        \hat{\tau} = &\mathbb{P}_m \left( \frac{G\{1-\hat{\pi}(X)\}}{1-q} \left[ \frac{A\{Y-\hat{\mu}_1(X)\}}{\hat{\omega}_1(X)} - \frac{(1-A)\{Y-\hat{\mu}_0(X)\}}{\hat{\omega}_0(X)} \right] \right.\\
        &\left. \quad + \frac{(1-G)\{1-\hat{\pi}(X)\}\hat{r}_1(X) \cdot A[Y-\hat{\tilde{\mu}}_1(X)]}{1-q} \frac{1}{\hat{\omega}_1(X)} \right.\\
        &\left. \quad - \frac{(1-G)\{1-\hat{\pi}(X)\}\hat{r}_0(X) \cdot (1-A)[Y-\hat{\tilde{\mu}}_0(X)]}{1-q} \frac{1}{\hat{\omega}_0(X)} \right.\\
        &\left. \quad + \frac{1-G}{1-q} [\hat{\mu}_1(X)-\hat{\mu}_0(X)+\gamma^{\top}X] \right).
    \end{aligned}
\end{equation}

The full implementation is summarized in Algorithm~1 in the Supplementary Material.

We next establish consistency properties for the proposed estimators. Proposition~\ref{prop:consistency know linear coeffs}
shows that, under the stated nuisance convergence conditions, the estimators
converge in probability to the target parameter \(\tau\).

\begin{proposition}[Consistency when linear coefficients are known]\label{prop:consistency know linear coeffs}
Suppose the linear drift coefficient \(\gamma\) is known.
\begin{enumerate}
    \item Under Assumptions \ref{assumption: unconfounding RCT} and \ref{assumption: linear posterior drift}, suppose $\hat{\pi}(x)$, $\hat{e}_1(x)$, $\hat{\mu}_1(x)$, and $\hat{\mu}_0(x)$ converge in probability under $\|\cdot\|_{\infty}$ to limits $\pi^*_I(x)$, $e_{1,I}^*(x)$, $\mu_{1,I}^*(x)$, and $\mu_{0,I}^*(x)$, respectively, with $\mu_{a,I}^*(x)=\mu_a(x)$ for $a\in\{0,1\}$, then $\hat{\tau}$ is consistent in probability as $n,\tilde{n} \to \infty$.
    \item Under Assumptions \ref{assumption: unconfounding RCT}, \ref{assumption: linear posterior drift}, and \ref{assumption: unconfounding OS}, suppose $\hat{\pi}(x)$, $\hat{e}_1(x)$, $\hat{e}_0(x)$, $\hat{\mu}_1(x)$, $\hat{\mu}_0(x)$, $\hat{\tilde{\mu}}_1(x)$, $\hat{\tilde{\mu}}_0(x)$, $\hat{r}_1(x)$, and $\hat{r}_0(x)$ converge in probability under $\|\cdot\|_{\infty}$ to limits $\pi^*_{II}(x)$, $e_{1,II}^*(x)$, $e_{0,II}^*(x)$, $\mu_{1,II}^*(x)$, $\mu_{0,II}^*(x)$, $\tilde{\mu}_{1,II}^*(x)$, $\tilde{\mu}_{0,II}^*(x)$, $r_{1,II}^*(x)$, $r_{0,II}^*(x)$, with $\mu_{a,II}^*(x)=\mu_a(x)$ and $\tilde{\mu}_{a,II}^*(x)=\tilde{\mu}_a(x)$ for $a\in\{0,1\}$, then $\hat{\tau}$ is consistent in probability as $n,\tilde{n} \to \infty$.
\end{enumerate}
\end{proposition}

\subsection{Unknown linear drift under OS unconfoundedness}\label{sec: unconfounded, linear parameters unknown}

In this subsection, we remain in the auxiliary OS-unconfounded regime of
Assumption~\ref{assumption: unconfounding OS}, but no longer assume that the
linear drift coefficient is known. Under this regime, the OS data can be used to estimate $\tilde{\mu}_a(X)$. Thus, we can estimate the expected potential outcomes in both datasets, use linear regression to estimate $\gamma$, and plug it into the estimator shown in formula (\ref{equation: estimator with known linear drift, A1+A2+A3}). To keep nuisance estimation, drift estimation, and final estimation separate, we use the source-stratified three-way split described in Algorithm~2 in the Supplementary Material.

\begin{theorem}[Property of $\hat{\gamma}$]\label{concentration of gamma}
    Suppose Assumptions~\ref{assumption: unconfounding RCT},
\ref{assumption: linear posterior drift}, and
\ref{assumption: unconfounding OS} hold. Let $\mathbb{P}_X$ denote the marginal distribution of $X$ in $(X,Y,A,G)$. Assume:
    \begin{enumerate}
        \item $\mathbb{P}_X$ has bounded support; for any $1\le k$, $l\le d$, $\mathbb{E}|X_{1,k}X_{1,l}|<\infty$; and $\mathbb{E}(X_1X_1^{\top})$ is positive definite, where $X_1\sim \mathbb{P}_X$.
        \item There exists a positive sequence $\{r_n\}$ with $r_n\downarrow 0$ and a constant $c_1>0$ such that
        \[
        \mathbb{P}\{ |\hat{\Delta}(X)-\Delta(X)|>t \} \le \exp\!\left(-c_1\left(\frac{t}{r_n}\right)^2\right),
        \]
        where $\hat{\Delta}$ is estimated from $n$ i.i.d.\ samples $(X_i,Y_i,A_i,G_i=1)$, and $X\sim\mathbb{P}_X$ is independent of these samples.
        \item Analogously, the OS CATE estimator $\hat{\tilde{\Delta}}$, trained on $\tilde n$ i.i.d.\ samples with $G=0$, satisfies the same tail condition for an independent $X\sim\mathbb{P}_X$, with $\hat{\tilde{\Delta}}(X)$, $\tilde{\Delta}(X)$, $\tilde{r}_{\tilde n}$, and $c_2$ replacing $\hat{\Delta}(X)$, $\Delta(X)$, $r_n$, and $c_1$, respectively.
    \end{enumerate}
    Then, for $n+\tilde n$ sufficiently large,
    \[
    \mathbb{E}\|\hat{\gamma}-\gamma\|_2 \le \tilde{C}\,\bar{r}_{n,\tilde n}\sqrt{\log m},
    \]
    where $\bar{r}_{n,\tilde n}=\max\{r_n,\tilde{r}_{\tilde n}\}$, and $\tilde{C}$ depends only on the support bound of $\mathbb{P}_X$ and the minimum eigenvalue of $\mathbb{E}(XX^{\top})$.
\end{theorem}

Theorem \ref{concentration of gamma} gives a finite-sample concentration rate for the plug-in estimator of the linear drift coefficient. In particular, the bound quantifies how estimation errors in the two CATE regressions propagate to $\hat{\gamma}$, and shows that $\hat{\gamma}$ is well controlled under sub-Gaussian-type tail behavior.

\begin{proposition}\label{prop:consistency unknown linear}
    Assume the conditions of Theorem \ref{concentration of gamma}, and suppose $\bar{r}_{n,\tilde n}\,\sqrt{\log(m)}\to 0$ as $n,\tilde n\to\infty$. Then the estimator $\hat{\tau}$ is consistent for $\tau$ in probability.
\end{proposition}

Proposition \ref{prop:consistency unknown linear} shows that once $\gamma$ is estimated accurately enough, the overall ATE estimator remains consistent despite the additional first-stage estimation step. This result justifies the three-way sample-splitting construction in the preceding algorithm, which decouples nuisance estimation, drift estimation, and target estimation.

\section{Estimation of ATE with unknown bias parameters under hidden confounding}\label{sec: hidden confounding, linear parameters unknown}

We now turn to the main setting, where the OS treatment assignment may be
affected by hidden confounding and the linear drift coefficient is unknown. In
this subsection, we continue to work under the linear posterior-drift model in
Assumption~\ref{assumption: linear posterior drift}, so that
\(
\tilde{\Delta}(X)=\Delta(X)+\gamma^\top X.
\)
Motivated by the known-drift estimator in
equation~\eqref{equation: estimator with known linear drift, A1+A2}, we retain
the same plug-in structure for the RCT correction and the target-population
averaging term, but replace the known coefficient \(\gamma\) by an estimator.
Thus, the main task is to estimate the CATE-drift coefficient \(\gamma\) when the
OS sample may be hidden-confounded.

The difficulty is that, under possible OS confounding,
\(
\tilde\mu_a(X):=\mathbb E\{Y(a)\mid X,G=0\}
\)
is not identified by the observed regression
\(\mathbb E(Y\mid X,A=a,G=0)\). Hence the OS causal contrast
\(\tilde\mu_1(X)-\tilde\mu_0(X)\) cannot be estimated directly from standard OS
outcome regressions. We instead use the RCT CATE \(\mu_1(X)-\mu_0(X)\) as an
internally valid anchor and estimate the drift term \(\gamma^\top X\) by
comparing the RCT-anchored model with a confounding-robust OS component. To
account for hidden confounding, we construct this OS component through a
Rosenbaum-type uncertainty set centered at the ordinary OS propensity score
\(\hat e_0\), and estimate \(\gamma\) by a minimax criterion over this uncertainty
set.

We formalize this idea next. Estimating \(\gamma\) requires CATE information from both RCT and OS samples. The RCT component is available through $\hat{\mu}_1(X)-\hat{\mu}_0(X)$; the main challenge lies in constructing a confounding-robust OS component.
Define the generalized propensity score as \(\tilde{e}_0(X,Y(0),Y(1)):=\mathbb{E}\{A\mid X,Y(0),Y(1),G=0\}\), and consider the transformed outcome:
\[
Y^* =Y\cdot \frac{A-\tilde{e}_0(X,Y(0),Y(1))}{\tilde{e}_0(X,Y(0),Y(1))\cdot(1-\tilde{e}_0(X,Y(0),Y(1)))},
\]
we have the following identification result:

\begin{proposition}[Identification of CATE]\label{prop:identification result}
    \(
    \mathbb{E}[Y^*|X=x, G=0]=\tilde{\Delta}(x).
    \)
\end{proposition}

In practice, \(\tilde{e}_0\) is not directly observable. Following \citet{Kallus2021minimaxpolicyconfounding}, we first estimate the ordinary propensity score \(e_0\), construct an interval around \(e_0\), and then solve a minimax problem with objective \(\sup_{E\in \text{interval}}\{\text{loss}(Y_E\sim X)\}\), where \(Y_E\) replaces \(\tilde{e}_0\) in \(Y^*\) by a candidate value \(E\).

The interval construction is motivated by a Rosenbaum-type sensitivity model. Specifically, for a sensitivity parameter \(\Gamma \geq 1\), one may require the generalized propensity score to satisfy the odds-ratio bound
\[
\Gamma^{-1}\leq \frac{(1-e_0(X))\cdot \tilde{e}_0(X,Y(0),Y(1))}{e_0(X)\cdot(1-\tilde{e}_0(X,Y(0),Y(1)))}\leq \Gamma.
\]
We use this display only as motivation for the uncertainty set: it says that the hidden-confounding propensity \(\tilde e_0\) should remain in an odds-ratio neighborhood of the ordinary propensity score \(e_0\).

As in Section~\ref{sec: unconfounded, linear parameters unknown}, we use a source-stratified three-way split.
Using the nuisance-training fold, we estimate
\(\pi(x),e_1(x),e_0(x),\mu_1(x),\mu_0(x)\) and define the RCT-based CATE estimator as
\(
\hat{\Delta}(x)=\hat\mu_1(x)-\hat\mu_0(x).
\)
For each OS observation \(Z_i=(A_i,X_i,Y_i)\) in the drift-estimation fold, the above sensitivity model leads us to allow the reciprocal generalized propensity weight \(W_i\) to vary over the interval $[a_i^\Gamma,b_i^\Gamma]$, where
\(
a_i^\Gamma=1+\Gamma^{-1}\{1/\hat e_0(X_i)-1\}
\),
\(
b_i^\Gamma=1+\Gamma\{1/\hat e_0(X_i)-1\}
\). Given a candidate vector \(W=(W_1,\ldots,W_{\tilde n})\), define the transformed OS outcome
\[
(Y_W)_i=Y_i\cdot \frac{A_iW_i-1}{1-1/W_i}.
\]
The empirical uncertainty set is therefore
\(
\mathcal{D}_{\tilde n}^\Gamma
:=\{W\in\mathbb{R}^{\tilde n}:a_i^\Gamma\leq W_i\leq b_i^\Gamma,\; i=1,\ldots,\tilde n\}.
\)

We estimate the drift coefficient by comparing the transformed OS outcomes with the RCT-anchored model \(\hat{\Delta}(X_i)+\eta^{\top}X_i\), while guarding against the worst admissible choice of \(W\). For a fixed \(W\), define
\[
\hat L_{\tilde n}(\eta;W)
:=\frac{1}{\tilde n}\sum_{i=1}^{\tilde n}\{(Y_W)_i-(\hat{\Delta}(X_i)+\eta^{\top}X_i)\}^2, \quad \hat{\bar L}_{\tilde n}(\eta)
:=\sup_{W\in\mathcal{D}_{\tilde n}^\Gamma}\hat L_{\tilde n}(\eta;W).
\]

The linear drift coefficient is estimated by the minimax rule
\(
\hat{\bar\gamma}\in\argmin_{\Vert\eta\Vert_\infty\leq H}\hat{\bar L}_{\tilde n}(\eta).
\)
Finally, we plug \(\hat{\bar\gamma}\) into the estimator \(\hat{\tau}\) above to obtain the final ATE estimate. The full implementation is summarized in Algorithm~3 in the Supplementary Material.

\begin{remark}
One tempting alternative is to estimate $\tilde{\mu}_1(X)-\tilde{\mu}_0(X)$ from OS data alone, then optimize a worst-case objective over an interval centered at the ordinary propensity score $e_0(x)$. We do not take this route because the OS-only contrast can inherit substantial bias when hidden confounding is present. Instead, our procedure uses the RCT CATE as an unconfounded anchor and estimates only the drift term needed to reconcile the RCT and OS causal contrasts.
\end{remark}

Correspondingly, a natural definition for population-level objective is 
\[
\bar L(\eta)
=
\mathbb E\left[
\sup_{w\in[a^\Gamma(X),\,b^\Gamma(X)]}
\left\{
Y\frac{Aw-1}{1-1/w}
-
\bigl(\Delta(X)+\eta^\top X\bigr)
\right\}^2
\,\middle|\,G=0
\right], 
\]
where \(
a^\Gamma(x)=1+\Gamma^{-1}\{1/e_0(x)-1\}\),
\(
b^\Gamma(x)=1+\Gamma\{1/e_0(x)-1\}
\).

Because the OS data may have unobserved confounders, \(\hat{\bar{\gamma}}\) need not converge to the true linear drift coefficient \(\gamma\). Instead, our objective is minimax optimality: we show that \(\hat{\bar{\gamma}}\) asymptotically minimizes the population supremum risk.

To isolate the statistical argument, we first assume $e_0(X)$ is known. The uncertainty set in the supremum loss is then \(\mathcal{D}_{\tilde{n}}'^\Gamma=\{W\in\mathbb{R}^{\tilde n}:a_i'^\Gamma\le W_i\le b_i'^\Gamma\}\), where \(a_i'^\Gamma=1+\Gamma^{-1}(1/e_0(X_i)-1)\) and \(b_i'^\Gamma=1+\Gamma(1/e_0(X_i)-1)\). Define \(\hat{\bar L}_{\tilde n}'(\eta):=\sup_{W\in\mathcal{D}_{\tilde n}'^\Gamma}\hat L_{\tilde n}(\eta;W)\), and let \(\hat{\bar\gamma}'\in\argmin_{\|\eta\|_\infty\le H}\hat{\bar L}_{\tilde n}'(\eta)\).

The technical assumptions needed in the theorems are summarized below.

\begin{assumption}[Common conditions for sup-risk concentration]\label{assumption: sup risk concentration}
The first-stage estimator $\hat{\Delta}$ is constructed independently of the OS empirical process and satisfies, for every sufficiently small $\delta>0$,
\(
\mathbb{P}\left\{\|\hat{\Delta}-\Delta\|_{\infty,\mathcal X} \leq n^{-1/2}c(\delta)\right\}\geq 1-\delta
\)
for some finite function $c(\delta)$. In addition, there exist constants $0<c_1<c_2<1$, $0<D, B_Y<\infty$ such that the covariate support $\mathcal{X}\subset[-D,D]^d$, $c_1\le e_0(x)\le c_2$ on the covariate support, and the domain of potential outcomes satisfying $\mathcal{Y}\subset [-B_Y,B_Y]$ for $a=0,1$ almost surely. Finally, $\Delta$ and $\hat{\Delta}$ are assumed to be uniformly bounded by a constant $M>0$ on $\mathcal X$.
\end{assumption}

\begin{theorem}[Concentration for the rectangular sup-risk]\label{concentrationofsupnorm}
Suppose Assumptions \ref{assumption: unconfounding RCT}, \ref{assumption: linear posterior drift}, \ref{assumption: sup risk concentration} hold and $H<\infty$. Then, for any given $\Gamma$, and for any sufficiently small $\delta>0$, with probability at least $1-\delta$, 
\[
\sup\limits_{\Vert\eta\Vert_{\infty} \leq H}|\hat{\bar{L}}'_{\tilde{n}}(\eta)-\bar{L}(\eta)| \leq C\cdot \bigg(\frac{1}{\sqrt{\tilde{n}}}\sqrt{\log(\frac{1}{\delta})}+ \frac{1}{\sqrt{n}}c(\delta)\bigg),
\]
where $C$ is a constant depending only on the constants in Assumption \ref{assumption: sup risk concentration}, $H$ and $\Gamma$.
\end{theorem}

Theorem \ref{concentrationofsupnorm} establishes a uniform concentration bound for the empirical minimax objective over the admissible parameter set. This result is the key technical bridge between sample-level optimization and population-level risk control.

\begin{corollary}\label{cor: concentration}
    Under the conditions of Theorem \ref{concentrationofsupnorm}, for any sufficiently small $\delta>0$, with probability at least $1-\delta$,
    \[
    \bar{L}(\hat{\bar{\gamma}}') -\inf\limits_{\Vert\eta\Vert_{\infty} \leq H} \bar{L}(\eta)\leq  2C\cdot\bigg(\frac{1}{\sqrt{\tilde{n}}}\sqrt{\log(\frac{1}{\delta})}+ \frac{1}{\sqrt{n}}c(\delta)\bigg).
    \]
\end{corollary}

Corollary 1 immediately implies near-optimality: the population sup-risk achieved by $\hat{\bar{\gamma}}'$ is close to the best achievable risk in the constrained class, up to the same stochastic error rate.

The preceding results are stated under a known ordinary propensity score. In practice, $e_0$ is unknown, and optimization is carried out over $\mathcal{D}_{\tilde n}^\Gamma$ rather than $\mathcal{D}_{\tilde n}'^\Gamma$. The next proposition quantifies the additional approximation error induced by using estimated propensity scores.

\begin{proposition}\label{prop: true and estimated propensity}
Let $h_i:=\hat{\Delta}(X_i)+\eta^{\top}X_i$. Then
\[
|\hat{\bar{L}}_{\tilde{n}}(\eta)-\hat{\bar{L}}_{\tilde{n}}'(\eta)|\leq 
\frac{2\Gamma}{\tilde{n}}\sum_{i=1}^{\tilde{n}}C_i(\eta)\cdot |\frac{1}{\hat{e}(X_i)}-\frac{1}{e(X_i)}|, 
\]
where 
\[
C_i(\eta) \leq 
\begin{cases}
    \displaystyle \max\limits_{W_i\in\{a_i^\Gamma\wedge a_i'^\Gamma, b_i^\Gamma\vee b_i'^\Gamma\}} |Y_i|\cdot|Y_iW_i-h_i| &\text{if}\; A_i = 1 \\[10pt]
    \displaystyle \max\limits_{W_i\in\{a_i^\Gamma\wedge a_i'^\Gamma, b_i^\Gamma\vee b_i'^\Gamma\}} \frac{|Y_i|}{(a_i^\Gamma\wedge a_i'^\Gamma-1)^3}\cdot|(Y_i+h_i)W_i-h_i| &\text{if}\;
    A_i = 0
\end{cases}
\]
\end{proposition}

Under additional regularity conditions (e.g., bounded $\|\eta\|_\infty$, bounded $Y_i$ and $X_i$, bounded $\hat{\Delta}(x)$, and strict positivity of the propensity score), the terms $C_i(\eta)$ are uniformly controlled. Consequently, the gap between $\hat{\bar L}_{\tilde n}(\eta)$ and $\hat{\bar L}_{\tilde n}'(\eta)$ is driven by propensity-score estimation error, which can be bounded via standard finite-sample guarantees. Combining these bounds with triangle inequalities yields near-optimality of $\hat{\bar\gamma}$ for the population worst-case loss $\bar L$.

\subsection{A general parametric setting}\label{sec:parametric-drift}
So far, we have focused on the linear posterior-drift specification. This choice is convenient for estimation and explanation, but in practice the discrepancy between the RCT and OS populations may follow a richer nonlinear pattern. To improve flexibility while keeping the model interpretable, we now consider a general parametric drift, which is assumed to be in a finite-dimensional function class.

\begin{assumption}[Parametric posterior drift]\label{assumption: parametric posterior drift}
$\tilde{\mu}_0(X) = \mu_0(X) + s(X)$ and $\tilde{\mu}_1(X) = \mu_1(X) + t(X)$, where $s(x), t(x) \in \{h_\eta:\eta\in \mathbb{R}^{d_1}\}$ and this function class is a vector space, which may include nonlinear transformations, interactions, or other scientifically motivated basis functions. 
\end{assumption}

Under Assumption \ref{assumption: parametric posterior drift}, 
the unknown object is no longer a linear coefficient
vector, but a drift function selected from a prespecified class. We estimate this function by the same uncertainty-set-based minimax principle used in the linear case, replacing the linear search space by the parametric class.

Using the notations of Theorem \ref{concentrationofsupnorm} and Assumption \ref{assumption: sup risk concentration}, we define the population and empirical worst-case risks for a generic drift function $h$ as follows:
\[
\bar L(h)
=
\mathbb E\left[
\sup_{w\in[a^\Gamma(X),\,b^\Gamma(X)]}
\left\{
Y\frac{Aw-1}{1-1/w}
-
\bigl(\Delta(X)+h(X))
\right\}^2
\,\middle|\,G=0
\right], 
\]
\[
    \hat{\bar{L}}_{\tilde{n}}(h) = \sup\limits_{W\in\mathcal{D}_{\tilde{n}}^\Gamma} \frac{1}{\tilde{n}} \sum_{i=1}^{\tilde{n}} ((Y_W)_i - (\hat{\Delta}(X_i)+h(X_i)))^2.
\]
In the parametric class, we take $h=h_\eta$, estimate the drift parameter by minimizing this empirical worst-case risk:
\[
\hat{\bar{\gamma}} \in \mathop{\text{argmin}}\limits_{\Vert\eta\Vert_{\infty} \leq H}\;\hat{\bar L}_{\tilde n}(h_\eta),
\]
and the final ATE estimator is obtained by inserting
it into the M-estimator. This
formulation emphasizes that the minimax procedure operates over a general
finite-dimensional drift class, rather than relying on the geometry of a linear working model.

As in Theorem~\ref{concentrationofsupnorm}, the ordinary propensity score is treated as known and the corresponding known-propensity version of worst-case loss function $\hat{\bar L}_{\tilde n}'(h_\eta)$ can be defined similarly. Since the proof strategy only relies on concentration and uniform control arguments that do not depend on linearity itself, Theorem \ref{concentrationofsupnorm} extends with minimal modification.

\begin{proposition}
    Suppose Assumptions \ref{assumption: unconfounding RCT}, \ref{assumption: sup risk concentration}, \ref{assumption: parametric posterior drift} hold, and the parameter of $t(x)-s(x)$ has $L_\infty$ norm at most $H$. Then, for any sufficiently small $\delta>0$, with probability at least $1-\delta$,
    \[
\sup\limits_{\Vert\eta\Vert_{\infty} \leq H}|\hat{\bar{L}}'_{\tilde{n}}(\eta)-\bar{L}(\eta)| \leq C\cdot \bigg(\frac{1}{\sqrt{\tilde{n}}}\sqrt{\log(\frac{1}{\delta})}+ \frac{1}{\sqrt{n}}c(\delta)\bigg),
\]
where $C$ is a constant depending only on the constants in Assumption \ref{assumption: sup risk concentration}, $H$ and $\Gamma$.
\end{proposition}

In finite samples, the minimax objective can be noisy, especially when the parametric class is moderately large. To stabilize optimization and reduce overfitting, we also consider a penalized version:
\(
    \hat{\bar{\gamma}} \in \mathop{\text{argmin}}\limits_{\Vert\eta\Vert_{\infty} \leq H}\;\bigg[\mathop{\text{sup}}\limits_{W\in \mathcal{D}_{\tilde{n}}^\Gamma}\frac{1}{\tilde{n}}\sum_{i=1}^{\tilde{n}}((Y_W)_i-(\hat{\Delta}(X_i)+h_\eta(X_i)))^2+\lambda\cdot q(\eta)\bigg],
\)
where $q(\eta)$ is a penalty term (e.g., $\|\eta\|_2^2$ or $\|\eta\|_1$), and $\lambda$ controls the strength of regularization. In applications, $\lambda$ can be selected by cross-validation or by a grid search.

\subsection{Nonparametric drifts}

We now further relax the modeling assumptions and allow posterior drift functions to be nonparametric. Specifically, we assume the drift functions belong to a H\"older class (see the Supplementary Material for the definition). To operationalize this model, we combine our minimax criterion with a sieve approach based on spline approximation \citep{wasserman2006nonparametric,Schumaker2007spline,vandervaart2023empiricalprocess}. This strategy provides a practical finite-dimensional surrogate for an infinite-dimensional optimization problem and leads to a risk bound analogous to Corollary \ref{cor: concentration}.

We first formalize the nonparametric drift condition. As above, we assume $\mathcal{X}$ is contained in $[-D,D]^d$ for some $D>0$.
\begin{assumption}[Nonparametric posterior drift]\label{assumption: nonparametric posterior drift}
$\tilde{\mu}_0(X) = \mu_0(X) + s(X)$ and $\tilde{\mu}_1(X) = \mu_1(X) + t(X)$, where $s,t\in C^{\alpha}(\mathcal{X})$ for some integer $\alpha>0$. 
\end{assumption}

The drift relevant for the conditional treatment effect is $h=t-s$. Under Assumption \ref{assumption: nonparametric posterior drift}, $h$ also belongs to $C^{\alpha}(\mathcal{X})$. We restrict the admissible drift class to the H\"older ball $C^{\alpha}(\mathcal{X})_{M_h}$, where $M_h$ controls the size of the class and can be selected based on prior domain knowledge or pilot-scale calibration.

To approximate this infinite-dimensional class, we use a tensor-product spline sieve $\mathcal{H}_N$. For a sieve level $N$, 
$\mathcal{H}_N$ is finite-dimensional, has dimension of order $N^d$, and satisfies the standard uniform spline approximation rate for H\"older functions. A more formal definition of these spline spaces is given in the Supplementary Material.

Using the worst-case risk notation introduced in Section~\ref{sec:parametric-drift}, our target and the corresponding sieve estimator are
\[
h_0 \in \mathop{\text{argmin}}\limits_{h\in C^{\alpha}(\mathcal{X})_{M_h}}  \bar{L}(h), 
\qquad \hat{h}_N \in \mathop{\text{argmin}}\limits_{h\in \mathcal{H}_N, \,\|h\|_{\infty}\leq aM_h} \hat{\bar{L}}_{\tilde{n}}(h), 
\]
where $a$ is a constant from \citet[Theorem~12.5]{Schumaker2007spline}, which can be taken as $((2\alpha+1)9^{\alpha})^d$. 
Intuitively, the sieve level $N$ controls the bias--variance trade-off: larger $N$ improves approximation accuracy but increases estimation variability. The next theorem
provides the balancing rate, and for simplicity, we treat the ordinary propensity score \(e_0\) as known in it.

\begin{theorem}[Concentration for the sup-norm in the nonparametric setting]\label{thm: nonparametric concentration}
Suppose Assumptions \ref{assumption: unconfounding RCT}, \ref{assumption: sup risk concentration}, \ref{assumption: nonparametric posterior drift} hold, and suppose the population minimizer $h_0$ exists. For the sieve estimator constructed above, take $N$ to be of order $\tilde{n}^{\frac{1}{d+2\alpha}}$. Then, for any sufficiently small $\delta>0$, with probability at least $1-\delta$,
\[
    \bar{L}(\hat{h}_N) - \bar{L}(h_0) \leq C_2\cdot\bigg(\frac{1}{\sqrt{\tilde{n}}}\sqrt{\log(\frac{1}{\delta})} + \frac{1}{\sqrt{n}}c(\delta) + \tilde{n}^{-\frac{\alpha}{d+2\alpha}}\bigg),
\]
where $C_2$ is a constant depending only on the constants in Assumption \ref{assumption: sup risk concentration}, $\alpha$, $d$, $M_h$ and $\Gamma$.
\end{theorem}

\begin{remark}
The estimator above relies on a user-specified smoothness level $\alpha$. In real applications, this smoothness is often unknown, so adaptive procedures (e.g., model-selection/cross-validation \citep{Birge1997modelselectionadaptive, BarronBirgeMassart1999, DingTarokhYang2018} or Lepski-type tuning \citep{Lepskii1992Asymptoticallyminimax,Birge2001Lepskimethod} across multiple sieve levels) may be preferable. Developing a fully adaptive version in the present minimax-transport framework is an interesting direction for future work.
\end{remark}

\section{Simulation Studies}\label{sec:simulations}
We conduct simulation studies to evaluate the finite-sample performance of the proposed minimax robust procedure when the drift is unknown and the observational study may be affected by hidden confounding. The experiments are designed to answer three questions: whether the method improves estimation of the target-population ATE $\tau$, whether it recovers the CATE-drift parameter $\gamma$, and how performance changes as $\Gamma$ varies. 

In the main text, we focus on a linear design that aligns with the working models used by the estimator. The data-generating mechanism builds on \citet{wu2024comparativeanalysisaveragetreatment} and \citet{Kallus2021minimaxpolicyconfounding}, while allowing both cross-source CATE drift and hidden confounding in the observational sample. Let $X$ denote the observed covariates, $U$ an unobserved confounder, and $Y(0),Y(1)$ the potential outcomes. We generate
\[
    X\sim N_6(0.2\cdot\mathbf{1}_6, 0.5\cdot\mathbf{I}_6),\quad
    U \sim N(0,0.5),\quad
    \epsilon_{11}, \epsilon_{10}\sim N(0,0.5),\quad
    \epsilon_{01}, \epsilon_{00}\sim N(0,0.8),
\]
\begin{align*}
    G=1: \, &Y(1) = 3+X^{\top}\theta +\epsilon_{11},\;Y(0) = 1-X^{\top}\theta+\epsilon_{10},\\
    G=0: \,&Y(1) = 3+X^{\top}\theta +X^{\top}\beta+2\cdot\texttt{u.strength}\cdot U+\epsilon_{01},\\
        &Y(0) = 1-X^{\top}\theta+X^{\top}\alpha+2\cdot\texttt{u.strength}\cdot U+ \epsilon_{00},
\end{align*}
where $\theta=(0.5,0.4,0.3,0.2,0.1,0.05)^{\top}$,
\(\alpha=\texttt{bias.strength}\cdot(0.3,0.6,-0.3,0,0.3,0.15)^{\top}\), \(\beta=\texttt{bias.strength}\cdot(1.2,0.3,0.6,0.9,-0.3,0.3)^{\top}\). 
Here, \texttt{bias.strength} controls the magnitude of the linear CATE drift and \texttt{u.strength} controls the strength of hidden confounding; all noise terms are mutually independent. Trial participation depends on both observed and unobserved variables through
\(
\Pr(G=1\mid X,U)=\operatorname{expit}(-1-0.2X^{\top}\theta-0.1U),
\)
where \(\operatorname{expit}(z)=\{1+\exp(-z)\}^{-1}\).
Treatment is randomized in the RCT with $\Pr(A=1\mid G=1,X,U)=0.5$. In the observational sample, treatment assignment is confounded by $U$:
\(
\Pr(A=1\mid G=0,X,U)
=\operatorname{expit}(-0.1+0.3X^{\top}\theta+4\cdot\texttt{u.strength}\cdot U).
\)

Estimation of $\gamma$ requires solving a finite-sample minimax problem. To improve numerical stability, we use two implementation refinements throughout the simulations. First, we add a quadratic penalty to the drift-estimation objective,
\(
\hat{\bar{\gamma}} \in
\operatorname*{argmin}_{\lVert\eta\rVert_{\infty} \leq H}
\bigg[\operatorname*{sup}_{W\in \mathcal{D}_{\tilde{n}}^\Gamma}
\frac{1}{\tilde{n}}\sum_{i=1}^{\tilde{n}}
\{(Y_W)_i-(\hat{\Delta}(X_i)+\eta^{\top}X_i)\}^2
+\lambda_{\mathrm{reg}}\lVert\eta\rVert_2^2\bigg].
\)
Second, while the inner supremum has a closed-form boundary solution (see the proof of Proposition \ref{prop: true and estimated propensity}), we solve the outer minimization with the Adam algorithm \citep{Kingma2014AdamAM}, which was stable across all sample sizes considered.

\begin{remark}
    Following \citet{Kallus2021minimaxpolicyconfounding}, the main design uses Gaussian covariates and Gaussian unobserved confounders. This choice is standard and widely used, but unbounded covariates can make the exact population-level interpretation of $\Gamma$ delicate. We therefore view the Gaussian design primarily as a comparison of algorithmic performance across \(\Gamma\) values, and we report bounded-support designs in the Supplementary Material to confirm that the qualitative conclusions are not driven by this issue.
\end{remark}

\begin{figure}[htbp]
     \centering
     \begin{subfigure}[b]{0.48\textwidth}
         \centering 
\includegraphics[width=\textwidth]{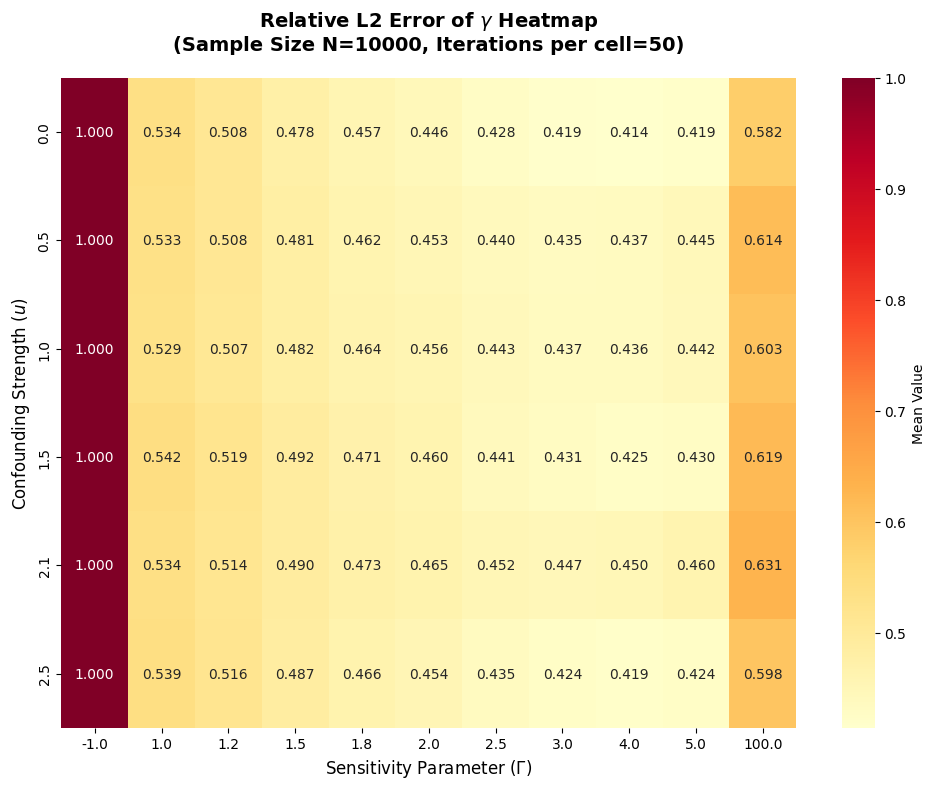}
         
     \end{subfigure}
     \hfill 
     \begin{subfigure}[b]{0.48\textwidth}
         \centering
\includegraphics[width=\textwidth]{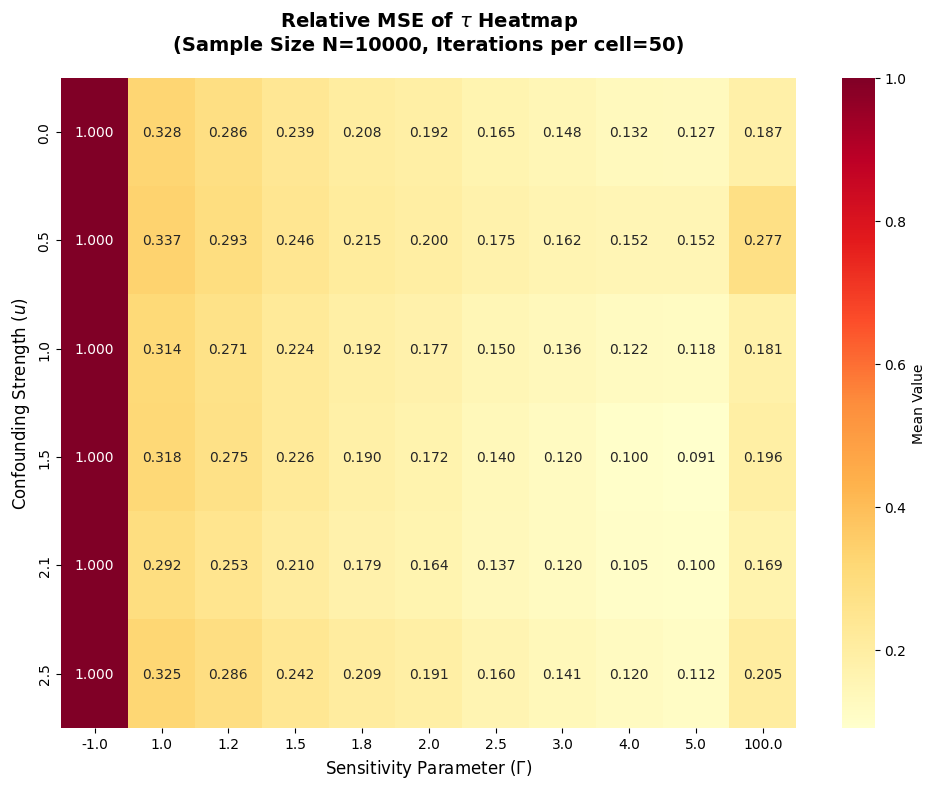}
         
     \end{subfigure}

     \vspace{2pt} 

     \begin{subfigure}[b]{0.48\textwidth}
        \centering
\includegraphics[width=\textwidth]{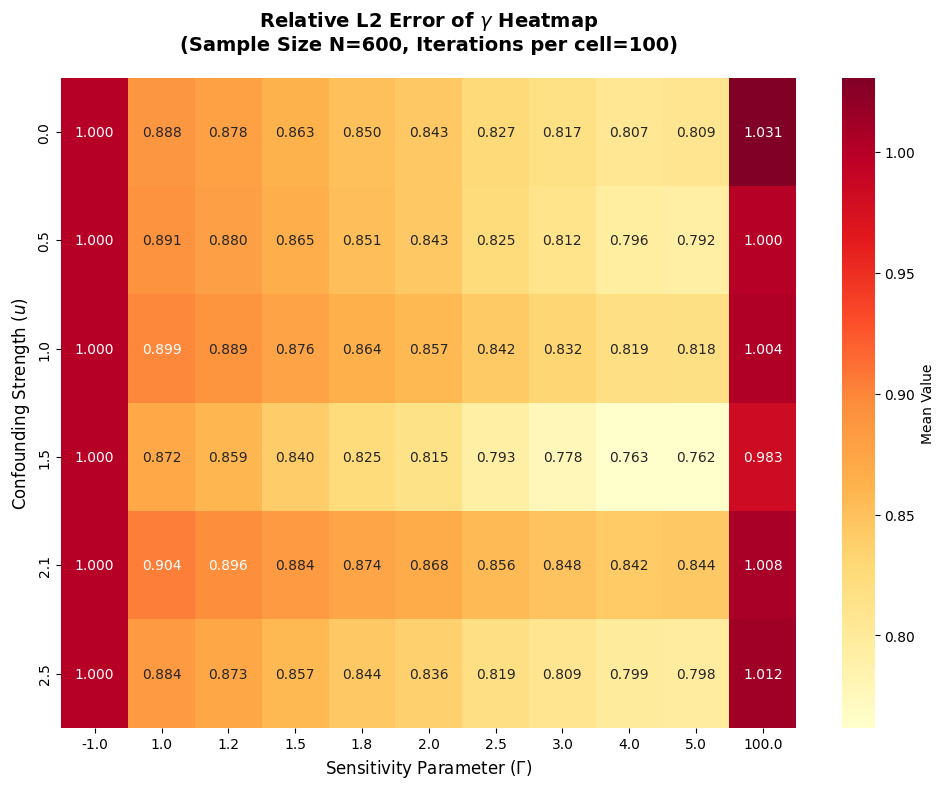}
        
     \end{subfigure}
     \hfill
     \begin{subfigure}[b]{0.48\textwidth}
        \centering
\includegraphics[width=\textwidth]{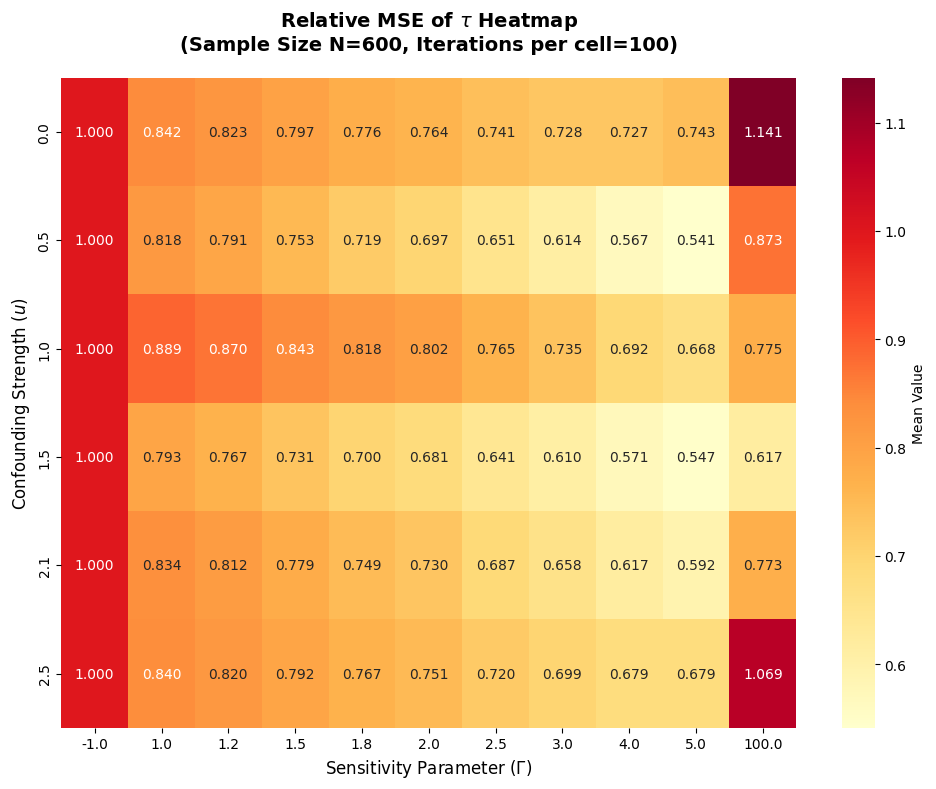}
        
     \end{subfigure}

     \caption{Linear design with moderate posterior drift: error ratios for estimating $\gamma$ (left) and $\tau$ (right) relative to the no-drift baseline $\Gamma=-1$. Lower values indicate better performance.}
     \label{fig:linear with moderate bias.strength}
\end{figure}

We consider four total sample sizes before sample splitting, $N_{\rm sim} \in \{10000,3000,1500,600\}$, and six hidden-confounding levels, $\texttt{u.strength} \in \{0.0,0.5,1.0,1.5,2.1,2.5\}$, where $\texttt{u.strength}=0.0$ corresponds to no unobserved confounding. The main experiment sets $\texttt{bias.strength}=1.5$, producing moderate posterior drift. We evaluate $\Gamma \in \{1.0,1.2,1.5,1.8,2.0,2.5,3.0,4.0,5.0,100.0\}$ and compare with a no-drift benchmark that sets $\gamma=0$; this benchmark is denoted by $\Gamma=-1$ in the figures. Figure \ref{fig:linear with moderate bias.strength} reports representative results for $N_{\rm sim}=10000$ and $600$, and the remaining sample sizes are shown in the Supplementary Material. Each scenario is replicated 50 or 100 times to reduce Monte Carlo variation.

Each heatmap reports an error ratio, defined as the estimation error of the proposed robust estimator at a given $\Gamma$ divided by the corresponding error of the no-drift benchmark. Rows correspond to hidden-confounding levels. The left panels report error ratios for $\gamma$, and the right panels report error ratios for the target ATE $\tau$. Across sample sizes, the proposed estimator typically improves on the no-drift benchmark, with larger gains in the larger-sample settings. The improvement is not limited to the case $\Gamma=1$, which allows drift but ignores hidden confounding; appropriately chosen values of $\Gamma$ yield additional gains.
    
The heatmaps also show a clear calibration pattern. As $\Gamma$ increases, errors for both $\gamma$ and $\tau$ usually decrease at first and then increase, producing a U-shaped relationship. This pattern is consistent with the role of $\Gamma$ as a sensitivity parameter: values that are too small under-adjust for hidden confounding, whereas values that are too large introduce unnecessary conservatism. The best performance is therefore attained at intermediate \(\Gamma\) levels.
    
    \begin{remark}
      We also repeat the main design without sample splitting; the corresponding results are reported in the Supplementary Material. The no-splitting implementation improves empirical performance uniformly across the scenarios considered. This finding is not surprising, because sample splitting is primarily used to support the theoretical analysis and can be conservative in finite samples. In applications with moderate sample sizes, using the full sample for all estimation stages may therefore provide better empirical efficiency.
    \end{remark}

The additional experiments in the Supplementary Material reinforce these conclusions across several departures from the main design. When the drift magnitude is reduced to one-fifth of the main setting, the no-drift benchmark becomes more competitive, but a properly calibrated \(\Gamma\) value still improves performance over the no-correction baseline for a broad range of confounding strengths. The bounded-support experiments show that the same qualitative behavior holds in designs where the population-level interpretation of $\Gamma$ is better controlled. Additional nonlinear-nuisance experiments indicate that the method remains useful under model misspecification and when nuisance functions are estimated with random forests. Finally, a quadratic-drift design, in which the CATE drift contains both linear and squared covariate terms, confirms that the proposed adjustment is not limited to purely linear drift patterns.

\FloatBarrier

\section{Case Study: Transporting ACTG 175 Evidence to the WIHS Cohort}\label{sec:case-study}

We study the practical performance of our framework by combining the ACTG 175 randomized trial with the WIHS cohort to estimate the ATE for the WIHS target population \citep{Dahabreh2023sensitivityanalysis}. 
This case study has three objectives. First, we examine whether
exact CATE transportability is empirically plausible. Second, we estimate the WIHS ATE using our linear-drift-adjusted procedure. Third, we investigate the finite-sample behavior of the method through a data-guided simulation that preserves the empirical covariate distribution.


\subsection{Data sources and study design}

AIDS Clinical Trials Group Study 175 (ACTG 175) is a randomized, double-blind, placebo-controlled trial comparing nucleoside monotherapy and combination therapy in adults with HIV-1 infection \citep{HammerKatzenstein1996ACTG}.
The trial started in December 1991 and enrolled 2,467 participants with screening CD4 counts between 200 and 500 cells/mm$^3$ (CD4 cells are a type of T helper cell that play a critical role in immune response, with a normal range of 500-1500 cells per microliter of blood). Participants were randomized to one of four regimens: zidovudine (AZT) alone, AZT+didanosine (ddI), AZT+zalcitabine (ddC), or ddI alone. Following \citet{Tsiatis2008covariateadjustment} and \citet{Dahabreh2023sensitivityanalysis}, we define the treatment contrast as $A=1$ for ddI monotherapy or either combination regimen, and $A=0$ for AZT monotherapy. The outcome in ACTG 175 is CD4 count measured at $20\pm5$ weeks after randomization.

Established in 1993, the Women's Interagency HIV Study (WIHS) is a large prospective cohort of women living with HIV or at elevated risk of infection \citep{Adimora2018WIHS}. Our target estimand is the average treatment effect (ATE) of $A=1$ versus $A=0$ in WIHS. In this study, the outcome is defined as the CD4 count at the next numbered visit, approximately six months after baseline, which serves as an observational analogue of short-term post-baseline immunologic response. Because treatment assignment in WIHS may be subject to unobserved confounding, we leverage the ACTG 175 randomized trial to obtain a more credible estimate of the ATE in the WIHS target population.

To combine ACTG 175 and WIHS, we made several design choices to improve comparability between the two data sources. In ACTG 175, we retained women only ($n=367$) to match the sex composition of the target population. In WIHS, we restricted attention to visits from 1994 to 1995, excluded pseudo-visit 0, and selected the first eligible visit for each participant as baseline. Baseline covariates include age, race, weight, and pre-treatment CD4 and CD8 counts, where CD8 cells are cytotoxic T cells that help eliminate infected cells. Treatment in WIHS was reconstructed from medication records: $A=0$ denotes AZT monotherapy, whereas $A=1$ denotes ddI monotherapy, AZT+ddI, or AZT+ddC. To improve overlap with ACTG 175, we further restricted WIHS participants to those with pre-treatment CD4 counts below 800 cells/mm$^3$, corresponding to the observed upper bound in ACTG 175. After excluding individuals with missing values, 429 WIHS participants remained.

For the primary analysis, we use age, race (represented by a binary indicator for non-White status), weight (kg), and pre-treatment CD4 and CD8 counts as baseline covariates. Pre-treatment CD4 and CD8 capture clinically important baseline immune status relevant to treatment assignment, subsequent immune response, and CATE-drift modeling. A secondary analysis using a restricted covariate set that excludes CD4 and CD8 is reported in Supplementary Section S5.3.

\subsection{Testing CATE Homogeneity Across ACTG 175 and WIHS}
Before fitting a posterior-drift model, we assess whether the CATE functions of the two datasets satisfy exact transportability. Using the notation introduced in Section~\ref{sec: setup}, the null hypothesis is
\[
H_0:\quad \tilde{\Delta}(X)=\Delta(X) \quad \text{a.s.}
\]
We apply the doubly robust procedure of \citet{armendariz2026testingeffecthomogeneityconfounding} to test this null hypothesis, and implementation details are provided in Supplementary Section S5.2.1. Because of the possible unobserved confounding of the WIHS data, rejection of this test may reflect posterior drift, unobserved confounding in WIHS, or both.

The transportability test uses the prespecified baseline covariates: age, race, weight, and pre-treatment CD4 and CD8 counts. We compare WIHS with three ACTG 175 reference samples: women only, men only, and the full ACTG 175 sample. Because the WIHS target population is female, the women-only comparison is the primary
one; the other two are included to check whether any evidence against CATE homogeneity
is specific to the sex-matched comparison or continues to appear under broader ACTG 175
reference samples.

Table~\ref{tab:cate_homogeneity_main} shows that
\(H_0\) is rejected for all three ACTG 175 reference
samples. Because this discrepancy may arise from posterior drift, unobserved confounding in WIHS, or both, the test does not identify either source separately and motivates our subsequent robust analysis, which allows for both. The restricted-covariate transportability test is reported in Supplementary Section S5.3.2.

\begin{table}[htbp]
\centering
\begin{tabular}{lrrrrrrc}
\toprule
ACTG sample & $n$ & Retained $n$ & Trimmed & $\hat{\theta}$ & SE & $p$-value & Reject \\
\midrule
Women & 796 & 751 & 5.65\% & 0.373 & 0.183 & 0.0412 & Yes \\
Men & 2198 & 1692 & 23.02\% & 0.467 & 0.152 & 0.0021 & Yes \\
All & 2565 & 2141 & 16.53\% & 0.449 & 0.122 & 0.000220 & Yes \\
\bottomrule
\end{tabular}
\captionsetup{singlelinecheck=false, justification=raggedright}
\caption{Results of transportability test between ACTG 175 and WIHS. The statistic $\hat{\theta}$ is reported on the standardized outcome scale, and the column ``Reject'' indicates whether the null hypothesis is rejected at the 5\% significance level.}
\label{tab:cate_homogeneity_main}
\end{table}

\subsection{Estimation strategy and findings}

Given the homogeneity-test results, we estimate the WIHS-target ATE using our proposed robust procedure, which combines trial and observational information while allowing residual hidden confounding through sensitivity parameter $\Gamma$. Given the moderate sample size, we use the no-splitting implementation to improve finite-sample efficiency. Both nuisance-learning strategies---linear models and random forests---use the primary baseline covariates. The main analysis uses a linear CATE-drift basis; Supplementary Section S5.2.2 reports analogous estimates using a quadratic drift basis. As in the simulation experiments, we also include a regularization penalty term in the loss function to improve estimation stability. All variables are scaled to $[0,1]$ during optimization and transformed back to the original CD4 scale for reporting.

For linear nuisance estimation, we use $H=1$, $\lambda_{\mathrm{reg}}=0.2$, learning rate $0.02$, and $400$ epochs. For random-forest nuisance estimation, we use $H=1$, tree depth $5$, learning rate $0.02$, $\lambda_{\mathrm{reg}}=0.5$, $350$ trees, minimum leaf size $15$, and $400$ epochs. Table~\ref{tab:tau_mean_std} reports both the benchmark estimates that ignore cross-source CATE differences and the confounding-adjusted estimates over $\Gamma$ values, with bootstrap standard deviations based on 1000 replications.

\begin{table}[htbp]
\centering
\begin{tabular}{lrrrr}
\toprule
Method & Baseline & $\Gamma = 1.0$ & $\Gamma = 1.2$ & $\Gamma = 1.5$ \\
\midrule
Linear
& 38.330 (16.007)
& 30.269 (16.558)
& 27.170 (17.289)
& 22.509 (18.503) \\
Random forest
& 36.117 (15.059)
& 35.755 (15.014)
& 34.746 (15.080)
& 33.224 (15.199) \\
\midrule
Method & $\Gamma = 2.0$ & $\Gamma = 3.0$ & $\Gamma = 5.0$ & \\
\midrule
Linear
& 14.695 (20.748)
& -0.685 (24.707)
& -30.737 (26.855) & \\
Random forest
& 30.689 (15.442)
& 25.666 (16.118)
& 15.014 (17.988) & \\
\bottomrule
\end{tabular}
\captionsetup{singlelinecheck=false, justification=raggedright}
\caption{Estimated WIHS ATE $\hat{\tau}$ under the baseline analysis that ignores cross-study CATE differences and under \(\Gamma\)-adjusted minimax analyses across $\Gamma$ values. Bootstrap standard deviations are reported in parentheses.}
\label{tab:tau_mean_std}
\end{table}

Table~\ref{tab:tau_mean_std} yields three main conclusions. First, under both nuisance-learning strategies, the estimates obtained after accounting for linear CATE differences are smaller than the baseline estimates that ignore those differences, which is consistent with the evidence against naive transportability. Second, under each strategy, the estimated treatment effect $\hat{\tau}$ decreases as $\Gamma$ increases, reflecting the increasingly conservative adjustment induced by stronger allowed hidden-confounding uncertainty. Third, except for the linear-nuisance analysis at very large $\Gamma$, the estimated effects remain positive under moderate sensitivity levels. Since prior clinical evidence favors active therapy \citep{HammerKatzenstein1996ACTG}, the negative values at $\Gamma=3$ and $5$ are best viewed as overly conservative scenarios rather than the most plausible conclusions for this application.

Overall, the case study suggests that analyses that ignore cross-source drift and hidden confounding can overstate the treatment effect when transporting trial evidence to WIHS. Our proposed procedure yields a more cautious and more interpretable assessment by explicitly modeling both concerns within a single framework.

\begin{remark}

Because the true WIHS ATE is not observed, we also conducted a data-guided
simulation calibrated to the women-only ACTG 175--WIHS sample; full details are reported in the Supplementary Material. 
The simulation fixes the empirical distribution of the primary baseline covariates and regenerates treatment and outcome using nuisance components estimated from the application. In both nuisance-learning settings, the MSE-minimizing sensitivity parameter is larger than one, and the estimated ATE decreases as \(\Gamma\) increases, matching the downward pattern as \(\Gamma\) increases in the real-data analysis. Restricted-covariate guided-simulation results are provided in the Supplementary Material. These results support the interpretation
that allowing nontrivial \(\Gamma\)-based robust adjustment can reduce the overstatement caused by indiscriminate exact-transportability analysis.

\end{remark}

\section{Conclusion}\label{sec:conclusion}

This paper develops a robust posterior-drift framework for transporting
randomized-trial evidence to an observational target population when exact
conditional-effect transportability may fail and the observational study may be
affected by hidden confounding. The central advancement is to treat cross-source
discrepancy as an unknown drift object that can be learned with the randomized
trial as an internally valid anchor, rather than imposing exact equality across
sources or specifying violations only through external bias functions. The method
therefore addresses a setting that is not covered by standard transportability
estimators, external-control borrowing procedures, or single-source robust analyses of hidden confounding.

The proposed estimator combines trial anchoring, observational target information,
and uncertainty-set calibration through a minimax criterion over a Rosenbaum-type uncertainty set. 
When the sensitivity parameter is calibrated to plausible degrees of hidden
confounding, the resulting estimator can mitigate the bias induced by naive
transportability assumptions and produce a more credible target-population effect estimate.
Simulations and the ACTG 175--WIHS application show that ignoring cross-source drift and hidden confounding can overstate transported effects,
whereas the proposed minimax robust  analysis gives more cautious and
interpretable estimates. Future work includes data-guided calibration of the
sensitivity parameter, adaptive choice of drift structure, and extensions to
longitudinal or multi-source settings.

\section*{Data Availability Statement}
ACTG 175 data are available at \url{https://archive.ics.uci.edu/dataset/890/aids%2Bclinical%2Btrials%2Bgroup%2Bstudy%2B175}; WIHS public-use data are available upon request at \url{https://www.niaid.nih.gov/research/wihs-public-dataset}
. Code is available from the authors upon request.

\section*{Acknowledgments}
ChatGPT 5.5/5.6 was used for language editing and manuscript preparation. All AI-assisted content was reviewed and verified by the authors.

\section*{Disclosure Statement}
The authors report there are no competing interests to declare.

\bibliography{references_jasa_rechecked}

@article{wu2024comparativeanalysisaveragetreatment,
  author  = {Wu, Peng and Luo, Shanshan and Geng, Zhi},
  title   = {On the Comparative Analysis of Average Treatment Effects Estimation via Data Combination},
  journal = {Journal of the American Statistical Association},
  volume  = {120},
  number  = {552},
  pages   = {2250--2261},
  year    = {2025}
}

@misc{sahoo2024learningbiasedsample,
  author = {Sahoo, Roshni and Lei, Lihua and Wager, Stefan},
  title  = {Learning from a Biased Sample},
  year   = {2022},
  note   = {arXiv preprint arXiv:2209.01754}
}

@article{Kallus2021minimaxpolicyconfounding,
  author  = {Kallus, Nathan and Zhou, Angela},
  title   = {Minimax-Optimal Policy Learning Under Unobserved Confounding},
  journal = {Management Science},
  volume  = {67},
  number  = {5},
  pages   = {2870--2890},
  year    = {2021}
}

@article{maity2021linearadjustmentbasedapproach,
  author  = {Maity, Subha and Dutta, Diptavo and Terhorst, Jonathan and Sun, Yuekai and Banerjee, Moulinath},
  title   = {A Linear Adjustment-Based Approach to Posterior Drift in Transfer Learning},
  journal = {Biometrika},
  volume  = {111},
  number  = {1},
  pages   = {31--50},
  year    = {2024}
}

@misc{cai2024transferlearningnonparametricregression,
  author = {Cai, T. Tony and Pu, Hongming},
  title  = {Transfer Learning for Nonparametric Regression: Non-Asymptotic Minimax Analysis and Adaptive Procedure},
  year   = {2024},
  note   = {arXiv preprint arXiv:2401.12272}
}

@article{tonycai2021nonparametricclassification,
  author  = {Cai, T. Tony and Wei, Hongji},
  title   = {Transfer Learning for Nonparametric Classification: Minimax Rate and Adaptive Classifier},
  journal = {The Annals of Statistics},
  volume  = {49},
  number  = {1},
  pages   = {100--128},
  year    = {2021}
}

@book{vandervaart2023empiricalprocess,
  author    = {van der Vaart, A. W. and Wellner, Jon A.},
  title     = {Weak Convergence and Empirical Processes: With Applications to Statistics},
  edition   = {2nd},
  publisher = {Springer},
  address   = {Cham},
  year      = {2023}
}

@book{wasserman2006nonparametric,
  author    = {Wasserman, Larry},
  title     = {All of Nonparametric Statistics},
  publisher = {Springer},
  address   = {New York},
  year      = {2006}
}

@article{neyman1923application,
  author  = {Neyman, Jerzy},
  title   = {On the Application of Probability Theory to Agricultural Experiments: Essay on Principles},
  journal = {Roczniki Nauk Rolniczych},
  volume  = {10},
  pages   = {1--51},
  year    = {1923}
}

@article{rubin1974estimating,
  author  = {Rubin, Donald B.},
  title   = {Estimating Causal Effects of Treatments in Randomized and Nonrandomized Studies},
  journal = {Journal of Educational Psychology},
  volume  = {66},
  number  = {5},
  pages   = {688--701},
  year    = {1974}
}

@inproceedings{Kingma2014AdamAM,
  author    = {Kingma, Diederik P. and Ba, Jimmy},
  title     = {{Adam}: A Method for Stochastic Optimization},
  booktitle = {International Conference on Learning Representations ({ICLR})},
  year      = {2015}
}

@article{Colnet2024causalcombiningsurvey,
  author  = {Colnet, B{\'e}n{\'e}dicte and Mayer, Imke and Chen, Guanhua and Dieng, Awa and Li, Ruohong and Varoquaux, Ga{\"e}l and Vert, Jean-Philippe and Josse, Julie and Yang, Shu},
  title   = {Causal Inference Methods for Combining Randomized Trials and Observational Studies: A Review},
  journal = {Statistical Science},
  volume  = {39},
  number  = {1},
  pages   = {165--191},
  year    = {2024}
}

@article{Degtiar2023generalizabilityreview,
  author  = {Degtiar, Irina and Rose, Sherri},
  title   = {A Review of Generalizability and Transportability},
  journal = {Annual Review of Statistics and Its Application},
  volume  = {10},
  pages   = {501--524},
  year    = {2023}
}

@inproceedings{Pearl2011tansportability,
  author    = {Pearl, Judea and Bareinboim, Elias},
  title     = {Transportability of Causal and Statistical Relations: A Formal Approach},
  booktitle = {Proceedings of the AAAI Conference on Artificial Intelligence},
  volume    = {25},
  pages     = {247--254},
  year      = {2011}
}

@article{Cole2010generalizing,
  author  = {Cole, Stephen R. and Stuart, Elizabeth A.},
  title   = {Generalizing Evidence from Randomized Clinical Trials to Target Populations: The {ACTG} 320 Trial},
  journal = {American Journal of Epidemiology},
  volume  = {172},
  number  = {1},
  pages   = {107--115},
  year    = {2010}
}

@article{Dahabreh2020Studydesignsfor,
  author  = {Dahabreh, Issa J. and Haneuse, Sebastien J.-P. A. and Robins, James M. and Robertson, Sarah E. and Buchanan, Ashley L. and Stuart, Elizabeth A. and Hern{\'a}n, Miguel A.},
  title   = {Study Designs for Extending Causal Inferences from a Randomized Trial to a Target Population},
  journal = {American Journal of Epidemiology},
  volume  = {190},
  number  = {8},
  pages   = {1632--1642},
  year    = {2021}
}

@article{Dahabreh2019ontherelation,
  author  = {Dahabreh, Issa J. and Robertson, Sarah E. and Hern{\'a}n, Miguel A.},
  title   = {On the Relation Between {G}-Formula and Inverse Probability Weighting Estimators for Generalizing Trial Results},
  journal = {Epidemiology},
  volume  = {30},
  number  = {6},
  pages   = {807--812},
  year    = {2019}
}

@article{Li2020improvingefficiency,
  author  = {Li, Xinyu and Miao, Wang and Lu, Fang and Zhou, Xiao-Hua},
  title   = {Improving Efficiency of Inference in Clinical Trials with External Control Data},
  journal = {Biometrics},
  volume  = {79},
  number  = {1},
  pages   = {394--403},
  year    = {2023}
}

@book{Schumaker2007spline,
  author    = {Schumaker, Larry L.},
  title     = {Spline Functions: Basic Theory},
  edition   = {3rd},
  series    = {Cambridge Mathematical Library},
  publisher = {Cambridge University Press},
  address   = {Cambridge},
  year      = {2007}
}

@article{Birge2001Lepskimethod,
  author  = {Birg{\'e}, Lucien},
  title   = {An Alternative Point of View on Lepski's Method},
  journal = {Lecture Notes--Monograph Series},
  volume  = {36},
  pages   = {113--133},
  year    = {2001}
}

@incollection{Birge1997modelselectionadaptive,
  author    = {Birg{\'e}, Lucien and Massart, Pascal},
  title     = {From Model Selection to Adaptive Estimation},
  booktitle = {Festschrift for Lucien Le Cam: Research Papers in Probability and Statistics},
  publisher = {Springer},
  address   = {New York},
  pages     = {55--87},
  year      = {1997}
}

@article{Dahabreh2023sensitivityanalysis,
  author  = {Dahabreh, Issa J. and Robins, James M. and Haneuse, Sebastien J.-P. A. and Saeed, Iman and Robertson, Sarah E. and Stuart, Elizabeth A. and Hern{\'a}n, Miguel A.},
  title   = {Sensitivity Analysis Using Bias Functions for Studies Extending Inferences from a Randomized Trial to a Target Population},
  journal = {Statistics in Medicine},
  volume  = {42},
  number  = {13},
  pages   = {2029--2043},
  year    = {2023}
}

@article{HammerKatzenstein1996ACTG,
  author  = {Hammer, Scott M. and Katzenstein, David A. and Hughes, Michael D. and Gundacker, Holly and Schooley, Robert T. and Haubrich, Richard H. and Henry, W. Keith and Lederman, Michael M. and Phair, John P. and Niu, Manette and Hirsch, Martin S. and Merigan, Thomas C.},
  title   = {A Trial Comparing Nucleoside Monotherapy with Combination Therapy in {HIV}-Infected Adults with {CD4} Cell Counts from 200 to 500 per Cubic Millimeter},
  journal = {New England Journal of Medicine},
  volume  = {335},
  number  = {15},
  pages   = {1081--1090},
  year    = {1996}
}

@article{Adimora2018WIHS,
  author  = {Adimora, Adaora A. and Ramirez, Catalina and Benning, Lorie and Greenblatt, Ruth M. and Kempf, Mirjam-Colette and Tien, Phyllis C. and Kassaye, Seble G. and Anastos, Kathryn and Cohen, Mardge and Minkoff, Howard and Wingood, Gina and Ofotokun, Igho and Fischl, Margaret A. and Gange, Stephen},
  title   = {Cohort Profile: The Women's Interagency {HIV} Study ({WIHS})},
  journal = {International Journal of Epidemiology},
  volume  = {47},
  number  = {2},
  pages   = {393--394i},
  year    = {2018}
}

@article{Tsiatis2008covariateadjustment,
  author  = {Tsiatis, Anastasios A. and Davidian, Marie and Zhang, Min and Lu, Xiaomin},
  title   = {Covariate Adjustment for Two-Sample Treatment Comparisons in Randomized Clinical Trials: A Principled Yet Flexible Approach},
  journal = {Statistics in Medicine},
  volume  = {27},
  number  = {23},
  pages   = {4658--4677},
  year    = {2008}
}

@misc{armendariz2026testingeffecthomogeneityconfounding,
  author = {Armendariz, Ana and Huber, Martin},
  title  = {Testing Effect Homogeneity and Confounding in High-Dimensional Experimental and Observational Studies},
  year   = {2026},
  note   = {arXiv preprint arXiv:2602.19703}
}

@article{Dahabreh2020Extendinginferences,
  author  = {Dahabreh, Issa J. and Robertson, Sarah E. and Steingrimsson, Jon A. and Stuart, Elizabeth A. and Hern{\'a}n, Miguel A.},
  title   = {Extending Inferences from a Randomized Trial to a New Target Population},
  journal = {Statistics in Medicine},
  volume  = {39},
  number  = {14},
  pages   = {1999--2014},
  year    = {2020}
}

@misc{yang2026improvingtreatmenteffectestimation,
  author = {Yang, Qinwei and Li, Jingyi and Wu, Peng and Yang, Shu},
  title  = {Improving Treatment Effect Estimation in Trials Through Adaptive Borrowing of External Controls},
  year   = {2026},
  note   = {arXiv preprint arXiv:2604.13973}
}

@article{Gao2025improvingrandomizedcontrolled,
  author  = {Gao, Chenyin and Yang, Shu and Shan, Mingyang and Ye, Wenyu and Lipkovich, Ilya and Faries, Douglas},
  title   = {Improving Randomized Controlled Trial Analysis via Data-Adaptive Borrowing},
  journal = {Biometrika},
  volume  = {112},
  number  = {2},
  pages   = {asae069},
  year    = {2025}
}

@article{Yang2023Elasticintegrative,
  author  = {Yang, Shu and Gao, Chenyin and Zeng, Donglin and Wang, Xiaofei},
  title   = {Elastic Integrative Analysis of Randomised Trial and Real-World Data for Treatment Heterogeneity Estimation},
  journal = {Journal of the Royal Statistical Society Series B: Statistical Methodology},
  volume  = {85},
  number  = {3},
  pages   = {575--596},
  year    = {2023}
}

@article{Dahabreh2024review,
  author  = {Dahabreh, Issa J. and Matthews, Anthony and Steingrimsson, Jon A. and Scharfstein, Daniel O. and Stuart, Elizabeth A.},
  title   = {Using Trial and Observational Data to Assess Effectiveness: Trial Emulation, Transportability, Benchmarking, and Joint Analysis},
  journal = {Epidemiologic Reviews},
  volume  = {46},
  number  = {1},
  pages   = {1--16},
  year    = {2024}
}

@article{Lepskii1992Asymptoticallyminimax,
  author  = {Lepskii, O. V.},
  title   = {Asymptotically Minimax Adaptive Estimation. I: Upper Bounds. Optimally Adaptive Estimates},
  journal = {Theory of Probability and Its Applications},
  volume  = {36},
  number  = {4},
  pages   = {682--697},
  year    = {1992}
}

@article{Robins1994EstimationofRegressionCoefficients,
  author  = {Robins, James M. and Rotnitzky, Andrea and Zhao, Lue Ping},
  title   = {Estimation of Regression Coefficients When Some Regressors Are Not Always Observed},
  journal = {Journal of the American Statistical Association},
  volume  = {89},
  number  = {427},
  pages   = {846--866},
  year    = {1994}
}

@article{Tan2010Boundedefficientdoublyrobust,
  author  = {Tan, Zhiqiang},
  title   = {Bounded, Efficient and Doubly Robust Estimation with Inverse Weighting},
  journal = {Biometrika},
  volume  = {97},
  number  = {3},
  pages   = {661--682},
  year    = {2010}
}

@article{Chan2016Globallyefficientnonparametric,
  author  = {Chan, Kwun Chuen Gary and Yam, Sheung Chi Phillip and Zhang, Zheng},
  title   = {Globally Efficient Non-Parametric Inference of Average Treatment Effects by Empirical Balancing Calibration Weighting},
  journal = {Journal of the Royal Statistical Society Series B: Statistical Methodology},
  volume  = {78},
  number  = {3},
  pages   = {673--700},
  year    = {2016}
}

@article{yang2020combining,
  author  = {Yang, Shu and Ding, Peng},
  title   = {Combining Multiple Observational Data Sources to Estimate Causal Effects},
  journal = {Journal of the American Statistical Association},
  volume  = {115},
  number  = {531},
  pages   = {1540--1554},
  year    = {2020}
}

@article{Han2025Federatedaptive,
  author  = {Han, Larry and Hou, Jue and Cho, Kelly and Duan, Rui and Cai, Tianxi},
  title   = {Federated Adaptive Causal Estimation ({FACE}) of Target Treatment Effects},
  journal = {Journal of the American Statistical Association},
  volume  = {120},
  number  = {551},
  pages   = {1503--1516},
  year    = {2025}
}

@article{Dorn2025doublyvalid,
  author  = {Dorn, Jacob and Guo, Kevin and Kallus, Nathan},
  title   = {Doubly-Valid/Doubly-Sharp Sensitivity Analysis for Causal Inference with Unmeasured Confounding},
  journal = {Journal of the American Statistical Association},
  volume  = {120},
  number  = {549},
  pages   = {331--342},
  year    = {2025}
}

@article{BarronBirgeMassart1999,
  author  = {Barron, Andrew and Birg{\'e}, Lucien and Massart, Pascal},
  title   = {Risk bounds for model selection via penalization},
  journal = {Probability Theory and Related Fields},
  year    = {1999},
  volume  = {113},
  number  = {3},
  pages   = {301--413}
}

@article{DingTarokhYang2018,
  author  = {Ding, Jie and Tarokh, Vahid and Yang, Yuhong},
  title   = {Model Selection Techniques: An Overview},
  journal = {IEEE Signal Processing Magazine},
  year    = {2018},
  volume  = {35},
  number  = {6},
  pages   = {16--34}
}

\end{document}